\documentclass[runningheads,a4paper,11pt]{llncs}
\usepackage[text={6.5in,9in},centering]{geometry}

\usepackage[T1]{fontenc}
\usepackage[utf8]{inputenc}

\usepackage{amsmath, amssymb}
\usepackage[english]{babel}
\usepackage{xspace, url, verbatim, array, float}
\usepackage[mathscr]{eucal}
\usepackage{comment}
\usepackage[version=3]{mhchem}
\usepackage{color}
\usepackage{cite}

\usepackage{graphicx}
\graphicspath{{./graphics/}}

\usepackage{dcolumn}
\newcolumntype{d}[1]{D{.}{.}{#1}}
\makeatletter
\newcolumntype{b}[1]{>{\boldmath\DC@{.}{.}{#1}}c<{\DC@end}}
\makeatother
\newcolumntype{M}{D{.}{.}{6}} %% for masses
\newcolumntype{C}{>{$}c<{$}}  %% for math numbers
\newcolumntype{R}{>{$}r<{$}}  %% for math numbers

\DeclareMathAlphabet{\mathpzc}{OT1}{pzc}{m}{it}

\newcommand{\todo}[1]{\xspace{\bfseries\sffamily\textcolor{red}{[#1]}}\xspace}

\newcommand\abs[1]{\left\lvert#1\right\rvert}

\newcommand\Rset{\ensuremath{\mathbb{R}}}

\begin{document}

\title{Heuristic algorithms for the Maximum Colorful Subtree problem}

\titlerunning{Heuristic algorithms for the Maximum Colorful Subtree problem}

\author{Kai D\"uhrkop\inst{1} \and Marie A. Lataretu\inst{1} \and
  W. Timothy J. White\inst{1,2} \and Sebastian B\"ocker\inst{1,3}}

\institute{Chair for Bioinformatics, Friedrich-Schiller-University, Jena,
  Germany \and Berlin Institute of Health, Berlin, Germany \and Contact:
  \url{sebastian.boecker@uni-jena.de}}

\date{\today}

\maketitle

\begin{abstract}
In metabolomics, small molecules are structurally elucidated using tandem
mass spectrometry (MS/MS); this resulted in the computational Maximum
Colorful Subtree problem, which is NP-hard.  Unfortunately, data from a
single metabolite requires us to solve hundreds or thousands of instances of
this problem; and in a single Liquid Chromatography MS/MS run, hundreds or
thousands of metabolites are measured.

Here, we comprehensively evaluate the performance of several heuristic
algorithms for the problem against an exact algorithm.  We put particular
emphasis on whether a heuristic is able to rank candidates such that the
correct solution is ranked highly.  We propose this ``intermediate''
evaluation because evaluating the approximating quality of heuristics is
misleading: Even a slightly suboptimal solution can be structurally very
different from the true solution.  On the other hand, we cannot structurally
evaluate against the ground truth, as this is unknown.  We find that
one particular heuristic consistently ranks the correct solution in
a top position, allowing us to speed up computations about 100-fold.  We also
find that scores of the best heuristic solutions are very close to the
optimal score; in contrast, the structure of the solutions can deviate
significantly from the optimal structures.
\end{abstract}

%%%%%%%%%%%%%%%%%%%%%%%%%%%%%%%%%%%%%%%%%%%%%%%%%%%%%%%%%%%%%%%%%%%%%%%%%%%%%
%%%%  SECTION INTRODUCTION  %%%%%%%%%%%%%%%%%%%%%%%%%%%%%%%%%%%%%%%%%%%%%%%%%
%%%%%%%%%%%%%%%%%%%%%%%%%%%%%%%%%%%%%%%%%%%%%%%%%%%%%%%%%%%%%%%%%%%%%%%%%%%%%

\section{Introduction}

Metabolomics characterizes the collection of all metabolites in a biological
cell, tissue, organ or organism using high-throughput techniques.  Liquid
Chromatography Mass Spectrometry (LC-MS) is one of the predominant
experimental platforms for this task.  Today, a major challenge is to
determine the identities of the thousands of metabolites detected in one
LC-MS run.  This is also true for related fields such as natural products
research \cite{wang16sharing}, biomarker discovery, environmental science, or
food science.  Tandem mass spectrometry (MS/MS) is used to derive information
about the metabolites' structures.  Interpretation of the hundreds to
thousands of MS/MS spectra generated in a single LC-MS run remains a
bottleneck in the analytical pipeline~\cite{wang16sharing}.  MS/MS data is
usually searched against spectral libraries \cite{stein12mass}, but only a
small number of metabolites (around 2\,\%) can be identified in this
manner~\cite{dasilva15illuminating}.  Recently, computational methods have
been developed that do not search in spectral libraries but rather in
molecular structure databases \cite{brouard17magnitude-preserving,
  brouard16fast, allen16computational, ruttkies16metfrag, verdegem16improved,
  tsugawa16hydrogen, duehrkop15searching, allen15competitive,
  shen14metabolite, ridder13automatic}.  CSI:FingerID
\cite{duehrkop15searching} has won several competitions on the identification
of small molecules from MS/MS
data\footnote{\url{http://casmi-contest.org/2017/results.shtml}}
\cite{schymanski17critical}; the web service for CSI:FingerID currently
analyzes more than 2000 queries a day.  At the heart of CSI:FingerID and its
variants \cite{brouard17magnitude-preserving, brouard16fast} lies the
computation of fragmentation trees, as these can be readily analyzed by
kernel-based methods using multiple kernel learning \cite{shen14metabolite}.

Fragmentation trees were introduced in 2008 \cite{boecker08towards} and were
initially targeted at the identification of the molecular formulas of small
molecules; later, it was shown that the structure of fragmentation trees
contains valuable information for structural elucidation of the underlying
molecule \cite{rasche12identifying, rasche11computing}.  Computing an optimum
fragmentation tree leads to the \textsc{Maximum Colorful Subtree} problem
\cite{boecker08towards}.  Unfortunately, this problem is NP-hard and also
hard to approximate \cite{rauf13finding, fertin17algorithmic}.  Algorithms
exist to solve the problem either heuristically \cite{rauf13finding} or
exactly \cite{boecker08towards, rauf13finding, white15speedy}.  The problem
is a variant of the well-studied \textsc{Graph Motif} problem
\cite{dondi11complexity, lacroix06motif}.

Approximation algorithms are algorithms for (usually) NP-hard problems with
provable guarantees on the distance of the returned solution to the optimal
one.  But in bioinformatics research, the objective function is usually only
a ``crutch'' used to find the optimum structure, whereas the value of the
objective function has little or no meaning.  To this end, heuristics in
bioinformatics are designed to find solutions structurally similar to the
optimum solution or, even better, the biological ground truth.  This makes it
intrinsically difficult to evaluate the performance of these heuristics, as
we have to define a measure on the structural similarity between the
heuristic solution and the biological ground truth; furthermore, the
biological ground truth has to be known.

We will use an alternative way to evaluate the performance of a heuristic:
For many applications, one biological instance results in a multitude of
computational instances, corresponding to candidates or hypotheses; the score
of the computational problem is used to rank these hypotheses.  Although the
ground truth may not be known for the structure of the best solution, we may
have information regarding the correct candidate or hypothesis.  To this end,
we can evaluate a heuristic based on its ability to top-rank the correct
candidate.

We propose several heuristics for the \textsc{Maximum Colorful Subtree}
problem, and evaluate these heuristics with regards to their ranking quality.
We find that one particular heuristic allows us to quickly confine
the set of candidates (molecular formulas of the precursor molecule).  This
constitutes a filter, such that optimum solutions have to be sought only for
a (preferably small) subset of candidates.  We also evaluate whether the
structure of the constructed solutions is similar to the optimum solution.
% Finally, we showcase on one real-world LC-MS/MS dataset how much using the
% heuristics can speed up computations in practice.

%%%%%%%%%%%%%%%%%%%%%%%%%%%%%%%%%%%%%%%%%%%%%%%%%%%%%%%%%%%%%%%%%%%%%%%%%%%%%
%%%%  SECTION  %%%%%%%%%%%%%%%%%%%%%%%%%%%%%%%%%%%%%%%%%%%%%%%%%%%%%%%%%%%%%%
%%%%%%%%%%%%%%%%%%%%%%%%%%%%%%%%%%%%%%%%%%%%%%%%%%%%%%%%%%%%%%%%%%%%%%%%%%%%%

\section{The Maximum Colorful Subtree problem}
\label{sec:problem}

Let $G=(V,E)$ be a node-colored, rooted, directed acyclic graph (DAG) with
root $r \in V$ and edge
weights $w: E \to \Rset$.  Let $C(G)$ be the set of colors used in $G$, and
let $c(v) \in C(G)$ be the color assigned to node $v \in V$.
We will consider subtrees $T=(V_T,E_T)$ of $G$ rooted at $r$.
Let $C(T)$ be the set of colors used in $T$.  We say that
$T$ is \emph{colorful} if all of its nodes have different colors.

The \textsc{Maximum Colorful Subtree} problem asks to find a colorful
$r$-rooted subtree $T$ of $G$ of maximum weight, where $G$ is a DAG with node
colors, edge weights and root $r \in V$.  We may assume that $G$ is (weakly)
connected and that $r$ is the unique source of~$G$, as we can remove all
nodes from the graph which cannot be reached by a path from the root~$r$,
without changing the optimal solution.

We note that previous work on the problem also makes the assumption of a
single source, albeit usually implicitly \cite{rasche11computing,
  boecker08towards, rauf13finding, white15speedy}.  From an algorithmic
standpoint, problem variants with or without a given root are ``basically
equivalent'': Given an algorithm that does not assume a fixed root $r$, we
can solve an instance of the problem variant with root $r$ by introducing a
superroot $r^*$ connected solely to $r$, sufficiently large edge weight
$w(r^*, r)$ and a new color for $r^*$.  For the reverse direction, we solve
the problem for every $r \in V$, then choose the best solution.

But there are two peculiarities when computing fragmentation trees that are
different from the general problem, and that we will make use of here: First,
any DAG used for fragmentation tree computation is \emph{transitive}: That
is, $uv \in E$ and $vw \in E$ implies $uw \in E$.  Second, a
coloring $c:V \to C(G)$ of DAG $G=(V,E)$ is \emph{order-preserving} if there
is an ordering `$\prec$' on the colors $C(G)$ such that $c(u) \prec c(v)$
holds for every edge $uv$ of~$G$ \cite{fertin17algorithmic}.  Computing
fragmentation trees naturally results in order-preserving colors, as nodes
can be colored by the fragment mass that is responsible for this node, and
edges exist only between nodes from larger to smaller masses.  The
\textsc{Maximum Colorful Arborescence} problem \cite{fertin17algorithmic}
asks to find a colorful induced subtree $T$ of $G$ of maximum weight, where
$G$ is a DAG with order-preserving colors and edge weights. See
\cite{fertin17algorithmic} for numerous complexity results.  Here, we will
stick with the name ``\textsc{Maximum Colorful Subtree} problem'', but
nevertheless assume that the coloring is order-preserving, unless indicated
otherwise.

%%%%%%%%%%%%%%%%%%%%%%%%%%%%%%%%%%%%%%%%%%%%%%%%%%%%%%%%%%%%%%%%%%%%%%%%%%%%%
%%%%  SECTION  %%%%%%%%%%%%%%%%%%%%%%%%%%%%%%%%%%%%%%%%%%%%%%%%%%%%%%%%%%%%%%
%%%%%%%%%%%%%%%%%%%%%%%%%%%%%%%%%%%%%%%%%%%%%%%%%%%%%%%%%%%%%%%%%%%%%%%%%%%%%

\section{Heuristics for the Maximum Colorful Subtree problem}
\label{sec:heuristics}

\begin{figure}[tb]
  \centering
  \includegraphics[width=0.8\linewidth]{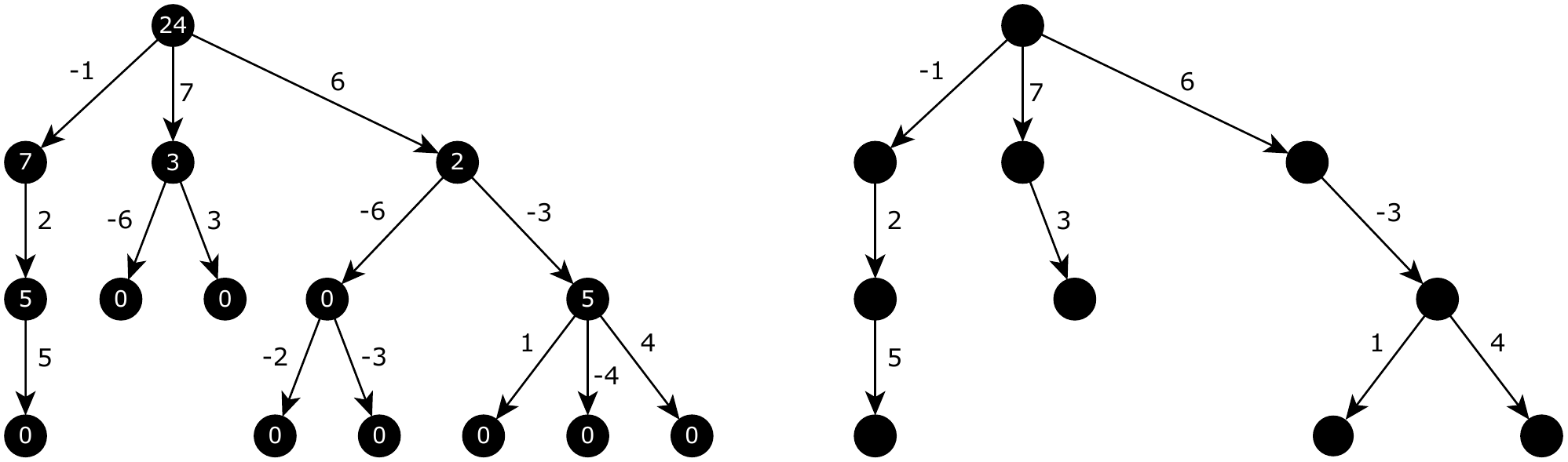}
  \caption{Illustration of the Remove Dangling Subtrees (RDS) postprocessing.
    Left: Input tree, where each node $v$ is labeled by its score $D[v]$.
    Right: Output tree of weight $24$.}
  \label{fig:rds}
\end{figure}

The following postprocessing methods can be applied to a tree $T=(V_T,E_T)$
\emph{after} any heuristic: The \textbf{Remove Dangling Edges} (RDE)
postprocessing iteratively removes edges $uv$ from $T$, where $v$ is a leaf
and $w(uv)<0$; this is repeated until no more such edges are found.  In
contrast, the \textbf{Remove Dangling Subtrees} (RDS) postprocessing does not
consider a single edge at a time, but rather induced subtrees: Each node $u
\in V_T$ is scored by the maximum weight of any induced subtree rooted
in~$u$.  Score $D[u]$ can be computed using dynamic programming:
\[
  D[u] = \sum\nolimits_{uv \in E_T} \max \{0,w(u,v) + D[v]\}
\]
Clearly, $D[u] \ge 0$.  For each edge $uv$ with $w(u,v) + D[v] < 0$ we remove
$uv$ and the subtree below it.  Both postprocessings can be computed in
$O(\abs{V_T})$ time using a tree traversal, as every edge is considered once
and $\abs{E_T} = \abs{V_T}-1$.  Figure~\ref{fig:rds} shows an example of the
RDS postprocessing.

%%%%%%%%%%%%%%%%%%%%%%%%%%%%%%%%%%%%%%%%%%%%%%%%%%%%%%%%%%%%%%%%%%%%%%%%%%%%%
%%%%%%%%%%%%%%%%%%%%%%%%%%%%%%%%%%%%%%%%%%%%%%%%%%%%%%%%%%%%%%%%%%%%%%%%%%%%%

We now present heuristics for finding a colorful subtree with root $r$ in a
transitive DAG with order-preserving coloring and unique source $r$.
\begin{itemize}
\item \emph{Kruskal-style.}  This heuristic sorts all edges of the graph by
  decreasing edge weight, then iteratively adds edges from the sorted list,
  ensuring that the growing subgraph is colorful and that each node has at
  most one incoming edges.  Since $r$ is the unique source of $G$, and since
  $G$ is transitive, this will ultimately result in a colorful subtree
  of~$G$.  This heuristic is similar to Kruskal's algorithm for computing an
  optimum spanning tree; it was called ``greedy heuristic''
  in~\cite{boecker08towards}.

\item \emph{Prim-style.}  This heuristic progresses similar to Prim's
  algorithm for calculating an optimal spanning tree: The tree $T=(V_T,E_T)$
  initially contains only the root $r$ of~$G$.  In every step, we consider
  all edges $uv$ with $u \in V_T$ and $v \notin V_T$ such that $c(v) \notin
  C(V_T)$; among these, we choose the edge with maximum weight and add it to
  the tree.  We repeat until all colors in the graph are used in the tree.
  This will usually result in a different tree than the Kruskal-style
  heuristic, due to the colorfulness constraint.

\item \emph{Insertion.}  This heuristic is a modification of the ``insertion
  heuristic'' from~\cite{rauf13finding}.  We again start with a tree
  $T=(V_T,E_T)$ containing only the root $r$ of~$G$.  The heuristic greedily
  attaches nodes labeled with unused colors.  For every node $v$ with
  $c(v)=c'$ unused, and every node $u$ already part of the solution, we
  calculate how much we gain by attaching $v$ to $u$.  To calculate this
  gain, we take into account the score of the edge $uv$ as well as the
  possibility of rerouting other outgoing edges of $u$ through~$v$:
\[
  I(u,v) := w(uv) + \displaystyle \sum_{x \in V_{T}, w(vx) > w(ux)} \bigl(
  w(vx) - w(ux) \bigr)
\]
  where we assume $w(uv) = -\infty$ if $uv \notin E$.  The node with maximum
  gain is then attached to the partial solution, and edges are rerouted as
  required.  See \cite{rauf13finding} for details; different from there, we
  do not iterate over colors in some fixed order but instead, consider all
  unused colors in every step.

\item \emph{Top-down.}  The top-down heuristic \cite{boecker08towards} is
  also greedy, but adds paths beginning in the root to the partial solution.
  The partial solution initially contains only the root $r$ of~$G$.  The
  heuristic greedily constructs a path starting at the root which is added to
  the partial solution; the next node of the path is chosen so that it
  maximizes the weight of the added edge, simultaneously ensuring that the
  partial solution remains colorful and does not violate the tree property.
  If no such edge exists, the algorithm restarts at the root, and searches
  for another path.  It terminates if no edge at the root can be selected.
  In the resulting tree, all internal nodes but the root have exactly one
  child.  This heuristic extends even simpler heuristics that attach all
  nodes to the root, which have been in frequent use for molecular formula
  determination from MS/MS data.

\item \emph{Critical Path$^1$.}  Again, we iteratively build the subtree;
  initially, the partial solution $T$ contains only the root $r$ of~$G$.  The
  score $S[u]$ of a node $u \in V$ is the maximum weight of a path $p$ from
  $u$ to any node $v$, such that $C(p) \cap C(V_T) \subseteq \{c(u)\}$; that
  is, the path does not use nodes with colors already present in the tree,
  except for the color of the starting node.  We can compute $S[u]$ using the
  recurrence
\begin{equation} \label{equ:critical-path}
  S[u] = \max\nolimits_{uv \in E, c(v) \notin C(V_T)} \{0,\ S[v] + w(uv)\}
\end{equation}
  where we use that the coloring of $G$ is order-preserving, since in that
  case no two nodes of the path have the same color.  We further assume $\max
  \emptyset = 0$ when computing \eqref{equ:critical-path}.  We iterate over
  the ordered colors in reverse order, computing $S[u]$ for all nodes $u$ of
  the active color.  The \emph{critical path} $p$ of maximum score can be
  found by backtracing from the maximum entry $S[u]$ with $u \in V_T$.  We
  add $p$ to $T$, then iterate, recomputing $S$ for the new set of used
  colors $C(V_t)$.  See Figure~\ref{fig:critical-path} for an example.

%%%%%%%%%%%%%%%%%%%%%%%%%%%%%%%%%%%%%%%%%%%%%%%%%%%%%%%%%%%%%%%%%%%%%%%%%%%%%
\begin{figure}[tb]
  \centering
  \includegraphics[height=0.28\linewidth]{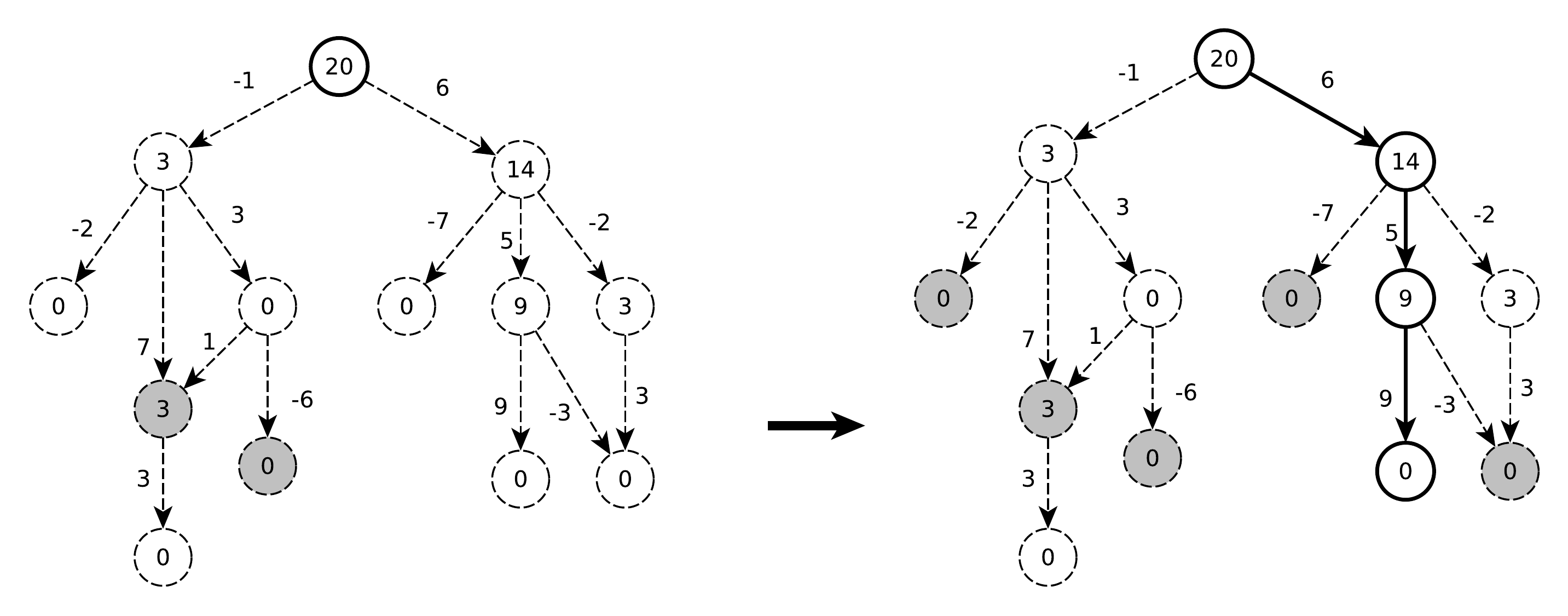}
  \hspace{0.07\linewidth}
  \includegraphics[height=0.28\linewidth]{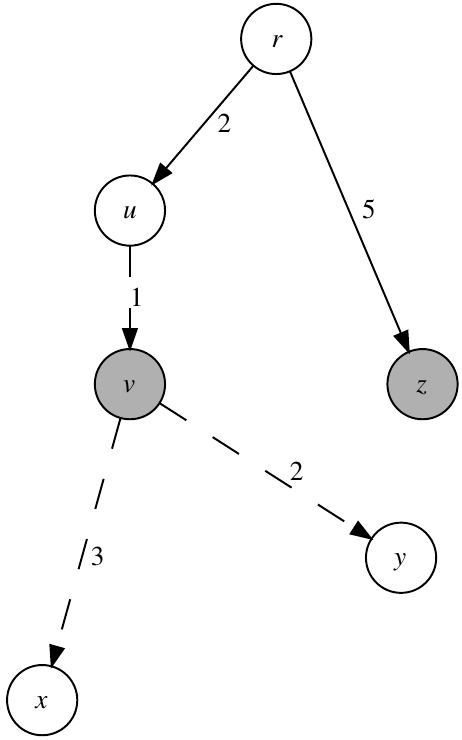}

  \caption{Left: Example for the \textit{Critical Path Heuristic}.  Nodes are
    labeled by score, solid lines show the tree, dashed lines the rest of the
    graph.  Grayed-out nodes have colors already used in the subtree.  Right:
    An example input graph for which Critical Path$^1$ produces a better tree
    than Critical Path$^2$.  Nodes $v$ and $z$ are the same color; all other
    nodes have distinct colors.  The two solid edges are the suboptimal tree
    output by Critical Path$^2$.  Critical Path$^1$ initially chooses the
    path $ruvx$ for a score of 6, then adds $vy$ for a total of~8.  Critical
    Path$^2$ begins in the same way by choosing the first edge $ru$ of the
    heaviest path for a score of 2, but in its second step it chooses the
    weight-5 edge $rz$, as the heaviest path starting with $uv$ has weight 4.
    No further edges can be added, so the total weight is~7.}
  \label{fig:critical-path}
\end{figure}
%%%%%%%%%%%%%%%%%%%%%%%%%%%%%%%%%%%%%%%%%%%%%%%%%%%%%%%%%%%%%%%%%%%%%%%%%%%%%

\item \emph{Critical Path$^2$.}  This heuristic also relies on critical
  paths, but adds, in each iteration, only the first edge of the critical
  path to the partial solution.  We note that this heuristic does not
  dominate the Critical Path heuristic, meaning that in certain cases, the
  subtree computed by this heuristic has smaller weight than that computed by
  the Critical Path heuristic; see Figure~\ref{fig:critical-path} (right) for
  an example.

\item \emph{Critical Path$^3$.}  This heuristic combines the Insertion heuristic
with the Critical Path heuristic: In each step the heuristic chooses the edge $uv$ with $u \in V_T$ that
maximizes the sum of critical path score and insertion score $S[u] + I(v)$.  
  % This heuristic uses Critical Path$^2$ to
  % find a good colorful subtree $T=(V_T,E_T)$.  It then builds a new tree on
  % the induced subgraph $G[V_T]$ by choosing, for each node, the incoming edge
  % with maximum weight.  Since $G[V_T]$ is a DAG where every node can be
  % reached from the root, this will again result in a tree rooted at~$r$.
  % Since the resulting tree has the same node set and the same number of
  % incoming edges to each node (namely, one) as the Critical Path$^2$ tree on
  % which it was based, it is easy to see that it never produces a worse
  % solution.

\item \emph{Maximum.}  All heuristics compute lower bounds of the maximum
  score; therefore the maximum score over all heuristic solutions is also
  a lower bound.
\end{itemize}

%%%%%%%%%%%%%%%%%%%%%%%%%%%%%%%%%%%%%%%%%%%%%%%%%%%%%%%%%%%%%%%%%%%%%%%%%%%%%

\paragraph{Time complexity of the heuristics.}

Let $n := \abs{V}$, $m := \abs{E}$, and $k := \abs{C(G)}$.  Clearly, $k \le
n$ and in applications, we usually have $k \ll n$.  Furthermore, $\abs{V_T}
\le k$ holds for the returned subtree $T=(V_T,E_T)$.
\begin{itemize}
\item It is easy to check that the Kruskal-style heuristic has time
  complexity $O(m \log n)$ for sorting all edges according to weight.
  Connectivity testing can then be performed in sub-logarithmic time per
  considered edge using a union-find data structure
  \cite{tarjan79class}; checking for colorfulness is easily accommodated by
  initially placing all nodes of the same color in the same component.
  The overall time complexity thus remains $O(m \log n)$.  Similarly, the
  Prim-style heuristic requires $O(m \log n)$ time.

\item For the Insertion heuristic, computing gain $I(u,v)$ for all $u$
  requires $O(k^2)$, since $u,x \in V_T$.  Hence, attaching one $v$ to the
  growing tree requires $O(k^2 n)$ time, resulting in $O(k^3 n)$ total
  running time. 

  But there exists a more complicated yet faster implementation for this
  heuristic: For each $v \in V$, we maintain two scores, $\mathit{in}(v)$ and
  $\mathit{out}(v)$, which correspond to the two terms on the RHS of the
  definition of $I(u, v)$.  Specifically, $\mathit{in}(v) = \max_{u \in V_T}
  w(uv)$, and $\mathit{out}(v) = \sum_{x \in V_T} \max \{0, w(vx) - w(p_T(x),
  x)\}$, where $p_T(x)$ is the parent of $x$ in $T$ for all $x \in T$.  To
  choose the next node to insert, we look for the node $v \in V$ maximizing
  $\mathit{in}(v) + \mathit{out}(v)$, ignoring nodes of already-used
  colors, which takes $O(n)$ time (and could in practice often finish early
  if we search in decreasing order of one of these terms, and know an upper
  bound on the other).  We then perform a single $O(k)$-time scan to find its
  optimal parent in the tree, and then perform two further updates: First, for
  all $u \in V$, set $\mathit{in}(u) \gets \max \{\mathit{in}(u), w(vu)\}$ and $\mathit{out}(u) \gets \mathit{out}(u) + w(uv) - w(p_T(v),v)$.
  Second, for all $x \in V_T$, check whether the incoming edge $yx$ (i.e., $y
  = p_T(x)$) can be improved by rerouting via $v$; if so, delete $yx$, insert
  $vx$ and for all $u \in V$ such that $w(yx) \le w(ux)$, set
  $\mathit{out}(u) \gets \max \{0, \mathit{out}(u) - (w(vx) - w(yx))\}$.  The
  second update needs $O(kn)$ time per inserted node, for $O(k^2n)$ overall.

\item The Top-down heuristic searches at most $k$ times for the maximum
  weight edge leaving a node; since there are $O(n)$ such edges, the running
  time is $O(k n)$.

\item For the Critical Path$^1$ heuristic, we need $O(m)$ time to compute
  values $S[u]$ and to identify the path of maximum weight.  This is repeated
  at most $k-1$ times, resulting in a total running time of $O(k m)$.  The
  same holds true for the Critical Path$^2$.  For Critical Path$^3$ we can
  again maintain an $\mathit{out}(v)$ table that contains the score bonus we
  get for attaching a node in the intermediate tree as child of $v$ and
  deleting its previous incoming edge.  After each insertion of an edge into
  the intermediate tree we have to perform the two update operations on
  $\mathit{out}$ which takes $O(kn)$ per insertion.  In total we need $O(k^2 n
  + k m)$ time to compute Critical Path$^3$.  In applications, $k$ is very
  small and $n \ll m$, so the $O(k m)$ part for calculating the critical
  paths requires most of the computation time.
%% FASTER?
\end{itemize}

%%%%%%%%%%%%%%%%%%%%%%%%%%%%%%%%%%%%%%%%%%%%%%%%%%%%%%%%%%%%%%%%%%%%%%%%%%%%%

\paragraph{Computing the $k$-best fragmentation trees exactly.}

Even if we do not trust the structural quality of the heuristic solution, the
above heuristics allow us to speed up fragmentation tree computation: We
first select a single candidate (molecular formula of the precursor) using
one of the heuristics, then compute the optimal solution for this instance
using an exact method \cite{boecker08towards, rauf13finding, white15speedy}.
In practice, this approach has two shortcomings: Even though certain
heuristics show a very good performance in selecting the correct molecular
formula (see below), this correct answer is not known to us in application;
but we will observe that the computed fragmentation tree will often not be
the optimum fragmentation tree, if we also consider other molecular formula
candidates and corresponding instances.

Even worse, it is usually not sufficient in application to select a single
best candidate using the heuristic, then re-compute the fragmentation tree
for the corresponding instance.  Instead, we usually want to know optimal
fragmentation trees for the $k$ best-scoring candidate.  This is independent
of whether results are reported to the user, who wants to use fragmentation
tree structure to survey if computations and, hence, the assigned precursor
molecular formula are trustworthy; or, if we perform some downstream
computational analysis based on fragmentation tree structure, such as
CSI:FingerID \cite{duehrkop15searching}.  In particular for ``large''
metabolites with mass beyond 600~Dalton, this is necessary because neither
the heuristics nor the exact method will always allow us to find the correct
candidate; only by considering several candidates, we can be sufficiently
sure that the correct answer is present.

We propose the following heuristic to compute optimum fragmentation trees for
the $k$-best molecular formula candidates: First, we compute heuristic
solutions for all candidates, and order molecular formula candidates
according to the heuristic score.  Next, we compute optimum fragmentation
trees for the $k$ best candidates; for small $k$, we can instead choose some
fixed parameter, such as 10 candidates.  We estimate the maximum $\Delta \ge
0$ of differences between the score of the optimum solution and the
corresponding heuristic solution, using those candidates where we know the
exact solution.  We now assume that the score difference is upper-bounded by
$\Delta$ for all candidates.  We continue to process candidates and compute
optimum fragmentation trees from the sorted list, updating the $k$-best
candidates and the corresponding score threshold; we stop computations when
the heuristic score of a candidate plus $\Delta$ is smaller than the current
score threshold.  Clearly, our assumption made above may be violated for
certain input, making this method a heuristic.

%%%%%%%%%%%%%%%%%%%%%%%%%%%%%%%%%%%%%%%%%%%%%%%%%%%%%%%%%%%%%%%%%%%%%%%%%%%%%
%%%%  DATA  %%%%%%%%%%%%%%%%%%%%%%%%%%%%%%%%%%%%%%%%%%%%%%%%%%%%%%%%%%%%%%%%%
%%%%%%%%%%%%%%%%%%%%%%%%%%%%%%%%%%%%%%%%%%%%%%%%%%%%%%%%%%%%%%%%%%%%%%%%%%%%%

\section{Data and Instances}

Details of how to transform the MS/MS spectrum of an (unknown) compound into
one or more instances of the \textsc{Maximum Colorful Subtree} problem have
been published elsewhere \cite{boecker08towards, boecker16fragmentation}; we
shortly recapitulate the process.  We consider all molecular formulas from
some ground set, such as, all molecular formulas from elements \ce{CHNOPS}.
We \emph{decompose} the precursor mass into all possible candidate molecular
formulas from this ground set; each candidate molecular formula corresponds
to one instance.  For each instance, we decompose the fragment peaks in the
MS/MS spectrum, ensuring that each fragment molecular formula is a subformula
of the candidate molecular formula for the precursor mass.  These molecular
formulas constitute the nodes of a graph; each node is colored by the peak it
stems from.  An edge is present between molecular formulas $u,v$ if and only
if $v$ is a 
proper subformula of $u$.  Now, both nodes and edges receive a certain weight
\cite{boecker16fragmentation}, based both on prior knowledge (e.g.,
distribution of loss masses) and the data (e.g., mass difference between a
peak and its hypothetical molecular formula); but as pointed in
\cite{boecker08towards}, we may assume that only edges are weighted.
Candidate molecular formulas of the precursor peak are ranked according to
the weight of the maximum colorful subtree in this graph.  SIRIUS 3.6 default
weights are used, see \cite{boecker16fragmentation}.

To evaluate whether a heuristic is capable of ranking the correct molecular
formula on the top position, we have to use reference data where the true
compound structure is known for each MS/MS measurement.  We use reference
compounds from GNPS~\cite{wang16sharing}; each reference compound is one
instance, corresponding to several graphs (for the different molecular
formula candidates of the precursor mass) we have to search in.  We then
filter instances: For example, we assume a mass accuracy of 10~ppm (parts per
million), and discard compounds where the precursor mass is missing or
outside outside this mass window.  All details can be found
in~\cite{boecker16fragmentation}.  This leaves us with 4\,050 compounds, each
of which is then transferred to usually many instances of the \textsc{Maximum
  Colorful Subtree} problem.  One reference compound resulted in between 1
and 21\,748 candidate molecular formulas, with median 53 and average 263.8
candidates.  To avoid proliferating running times, we consider only the 60
most intense peaks in a MS/MS spectrum that can be decomposed, which is again
SIRIUS 3.6 default behavior.  We fix the SIRIUS tree size parameter, which is
usually adapted at runtime, at $-0.5$.  In addition, we switch off SIRIUS'
spectral recalibration.

%%%%%%%%%%%%%%%%%%%%%%%%%%%%%%%%%%%%%%%%%%%%%%%%%%%%%%%%%%%%%%%%%%%%%%%%%%%%%
%%%%  RESULTS  %%%%%%%%%%%%%%%%%%%%%%%%%%%%%%%%%%%%%%%%%%%%%%%%%%%%%%%%%%%%%%
%%%%%%%%%%%%%%%%%%%%%%%%%%%%%%%%%%%%%%%%%%%%%%%%%%%%%%%%%%%%%%%%%%%%%%%%%%%%%

\section{Results}

We applied all but the Critical Path heuristics using the RDS postprocessing.
We do not evaluate the RDE postprocessing, as it is dominated by RDS (that
is, the score is at least as good, in all cases) which, in turn, dominates
solutions without postprocessing.  Furthermore, both postprocessings are very
fast in practice.  For the Critical Path heuristics, RDS cannot improve a
solution for variants 1 and~2; for variant~3 this is possible in principle,
but very unlikely.  To keep results of the Critical Path heuristics
consistent, we disabled the RDS postprocessing for variant~3, too.  All
heuristics were implemented in Java 8. 
For the exact method, we use the Integer Linear
Program (ILP) from \cite{rauf13finding} with the CPLEX solver 12.7.1 (IBM,
\url{https://www.ibm.com/products/ilog-cplex-optimization-studio}).

%%%%%%%%%%%%%%%%%%%%%%%%%%%%%%%%%%%%%%%%%%%%%%%%%%%%%%%%%%%%%%%%%%%%%%%%%%%%%
\begin{figure}[tb]
  \centering
  \includegraphics[width=0.48\linewidth]{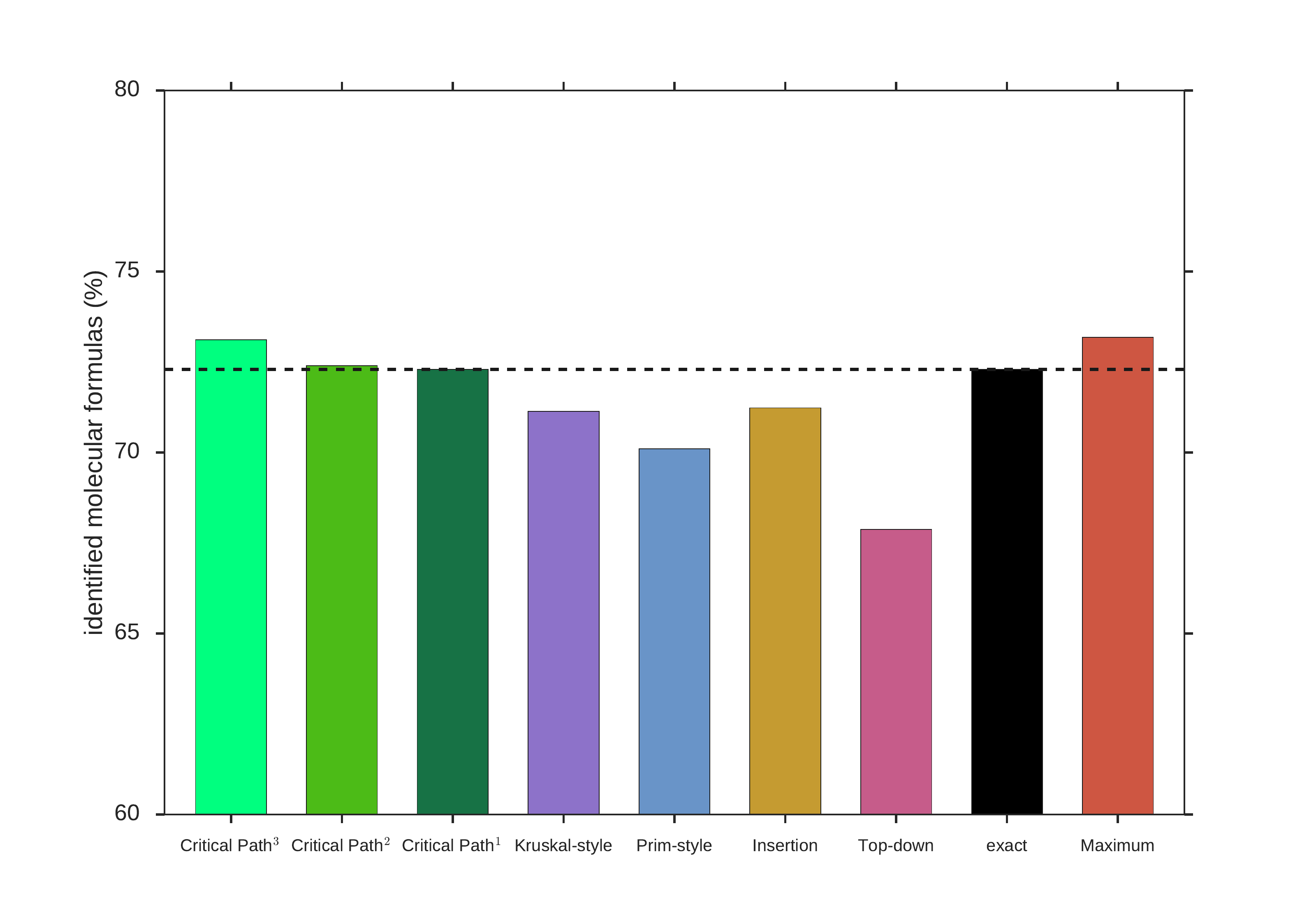}
  \hspace{0.02\linewidth}
  \includegraphics[width=0.48\linewidth]{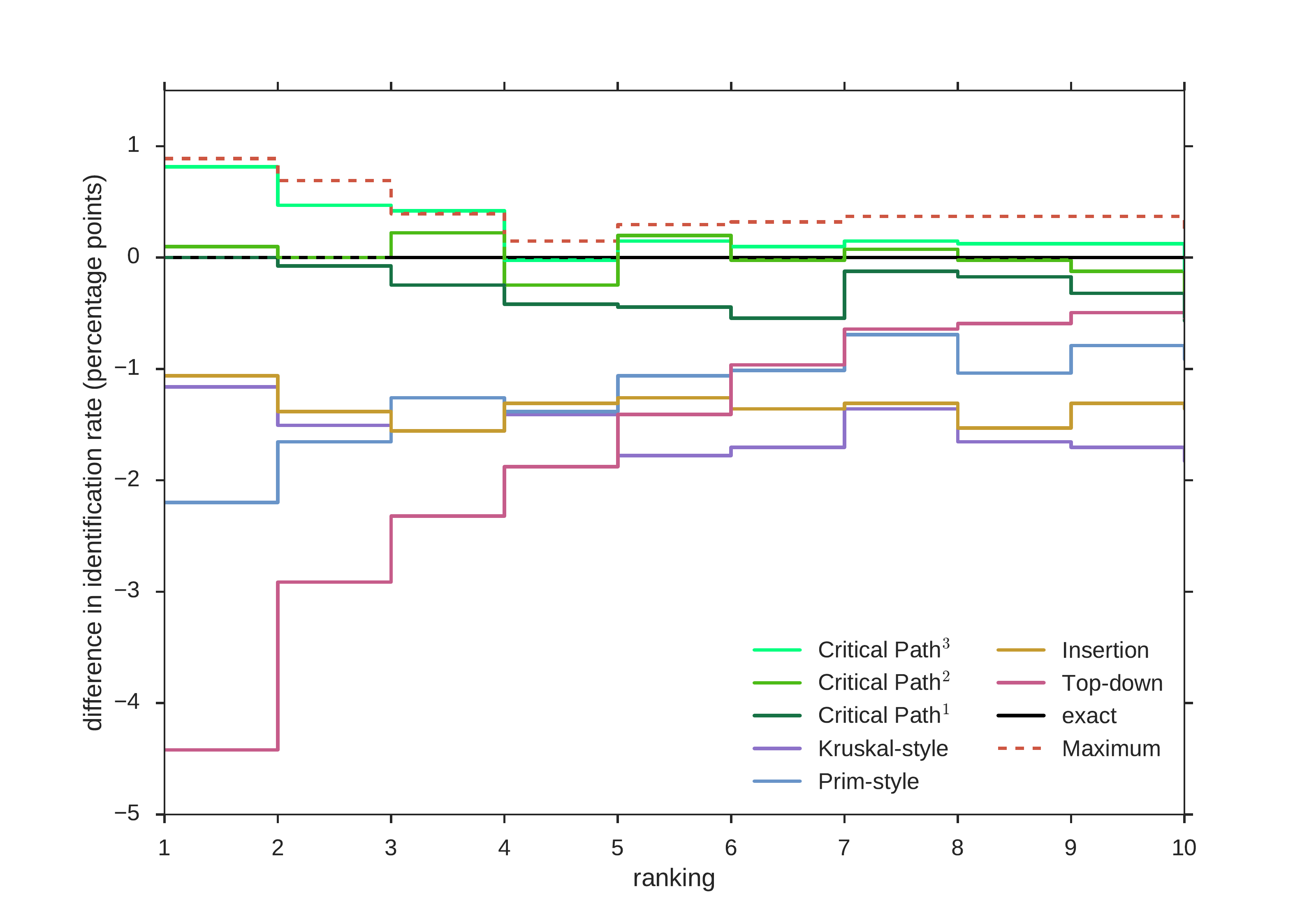}

  \caption{Performance evaluation, finding the correct molecular formula.
    Left: \emph{Percentage} of compounds where the correct molecular formula
    received the highest score.  Note the zoom of the y~axis.  Right:
    \emph{Percentage point difference} against exact computations; how often
    is the correct answer part of the top~$k$ output of each method?}
  \label{fig:ranking}
\end{figure}
%%%%%%%%%%%%%%%%%%%%%%%%%%%%%%%%%%%%%%%%%%%%%%%%%%%%%%%%%%%%%%%%%%%%%%%%%%%%%

First, we evaluated the power of the different heuristics to rank the correct
answer (molecular formula) at the top position; see Fig.~\ref{fig:ranking}
(left).  We also compared against the exact solution.  We observe similar
identification rates for the critical path heuristics, the maximum heuristic
and the exact method.  To test whether this trend is true not only for the
top rank, but also for the top~$k$ ranks, we also evaluated how often any
method is capable to rank the correct answer in its top~$k$, for varying $k$;
see Fig.~\ref{fig:ranking} (right).  Identification rates differ much
stronger when varying $k$ for one method than for different methods and
one~$k$; to this end, we normalize identification rates by subtracting the
identification rate of the exact method.  We see that all heuristics but
critical path result in inferior rankings, loosing one or more percentage
points for most ranks.  In contrast, the critical path heuristics rank
solutions with comparable power as the exact method, and the later two
variants often outperform the exact method.  Somewhat surprisingly, the
maximum over all heuristics performs even better than the best heuristic.
%This excellent performance is predominantly due to combining the Critical
%Path$^3$ and Insertion heuristics: If we take the maximum of only these two
%heuristics, we reach 2954 correct identifications, compared to 2956 correct
%identifications when considering all heuristics. 

%%%%%%%%%%%%%%%%%%%%%%%%%%%%%%%%%%%%%%%%%%%%%%%%%%%%%%%%%%%%%%%%%%%%%%%%%%%%%
\begin{figure}[tb]
  \centering
  \includegraphics[width=0.8\linewidth]{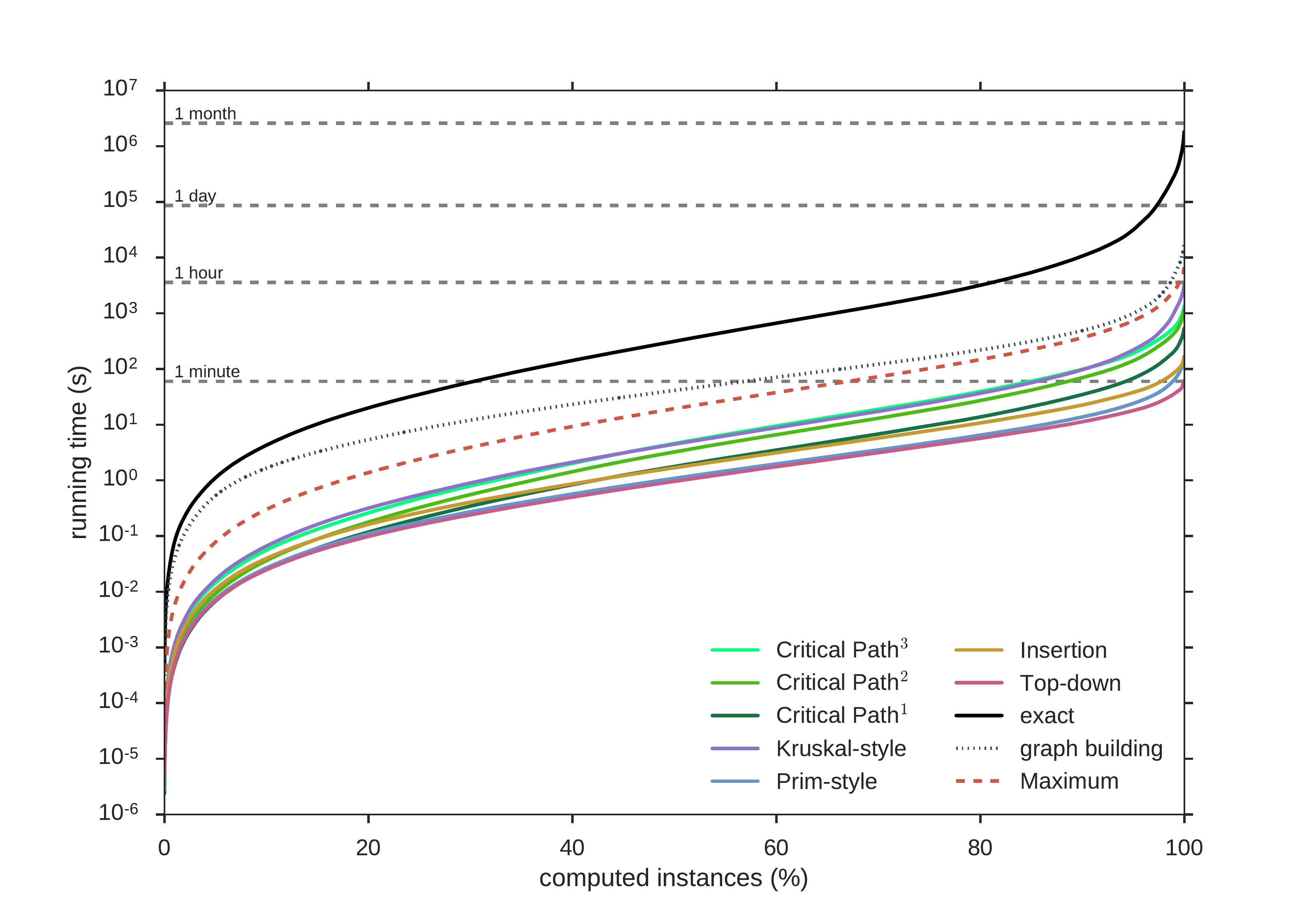}

  \caption{Running times of the different methods.  One instances consists of
    all graphs generated for one compound in the dataset, considering all
    decompositions of the precursor mass.  For each method, instances are
    sorted with respect to running time, and we report amortized running
    times.  We also report running times for constructing the instance DAGs
    and for the exact method.}
  \label{fig:running-times}
\end{figure}
%%%%%%%%%%%%%%%%%%%%%%%%%%%%%%%%%%%%%%%%%%%%%%%%%%%%%%%%%%%%%%%%%%%%%%%%%%%%%

Second, we compared running times of the different methods.  Running times
were measured using a single thread on an Intel E5-2630v3 at 2.40~GHz with
64~GB RAM.  The total running time for the exact methods over all instances
is almost one month, underlining the importance of speeding up computations.
But also note that solving all instances exactly requires only about 100-fold the
time required for constructing the instance graphs.  For each method, we
sorted all instances by running time; we then reported how much time is
required to solve, say, the 90\,\% ``easiest'' instances for that method.
Generally, this ordering is different for each method.  For all methods, we
observe that the ``hardest'' 5\,\% of the instances are responsible for most
of the total running time; this has been observed before \cite{rauf13finding,
  boecker16fragmentation}.  In comparison to the exact method, all heuristics
are very fast, and at least two orders of magnitude faster.  In particular,
\emph{each} heuristic is faster than the method for constructing the instance
graphs; running all heuristics, as required for the maximum heuristic,
requires about the same time as the graph construction.  Comparing
heuristics' running times, we see that the Kruskal-style heuristic is slowest;
%we see that the Insertion heuristic is slowest, as
%suggest by worst-case considerations; 
and that the first variant of the
Critical Path heuristic is faster in practice than variants 2 and~3, but not
significantly.

%%%%%%%%%%%%%%%%%%%%%%%%%%%%%%%%%%%%%%%%%%%%%%%%%%%%%%%%%%%%%%%%%%%%%%%%%%%%%
\begin{figure}[tb]
  \centering
  \includegraphics[height=0.39\linewidth]{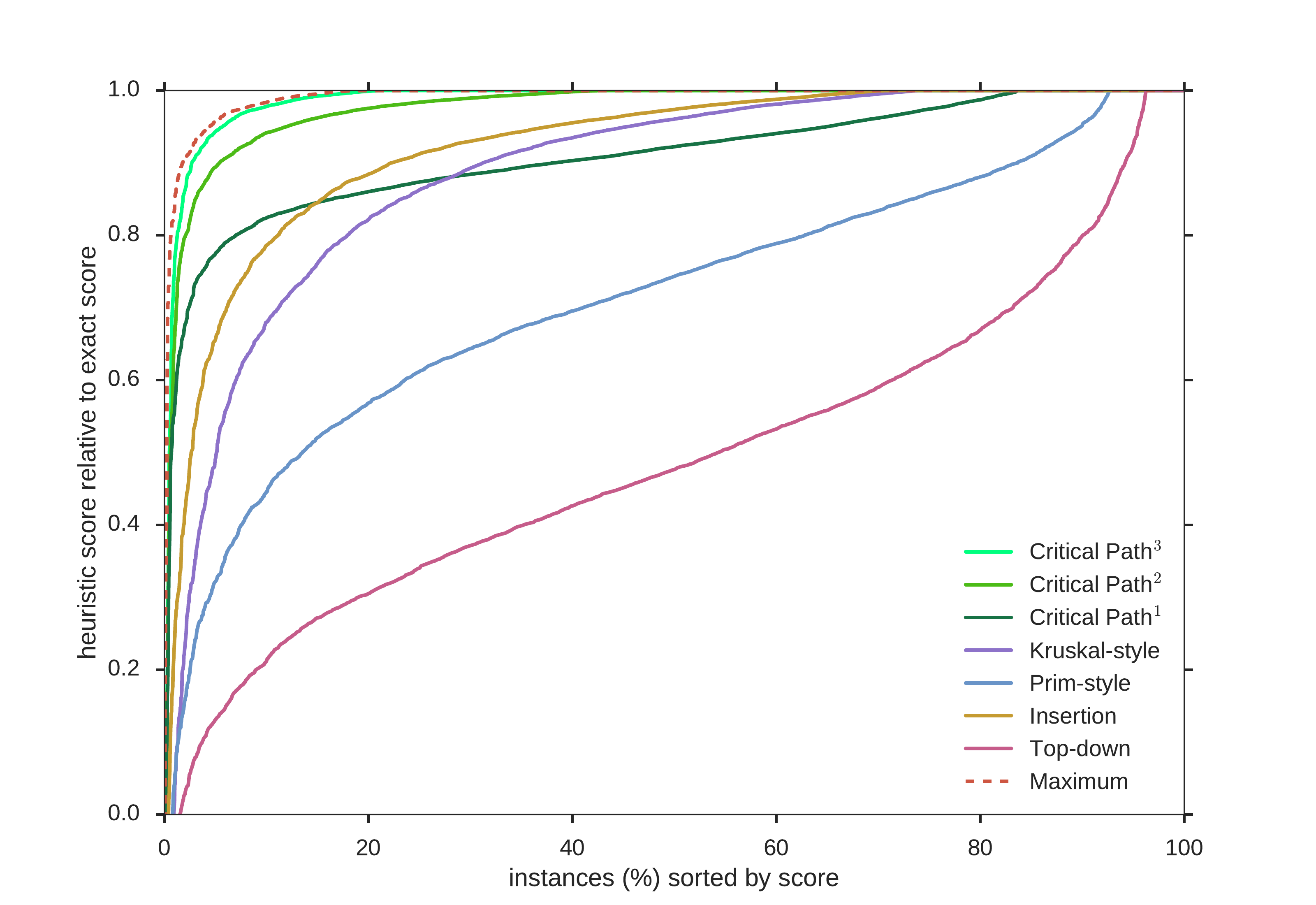}
  \hspace{0.02\linewidth}
  \includegraphics[height=0.39\linewidth]{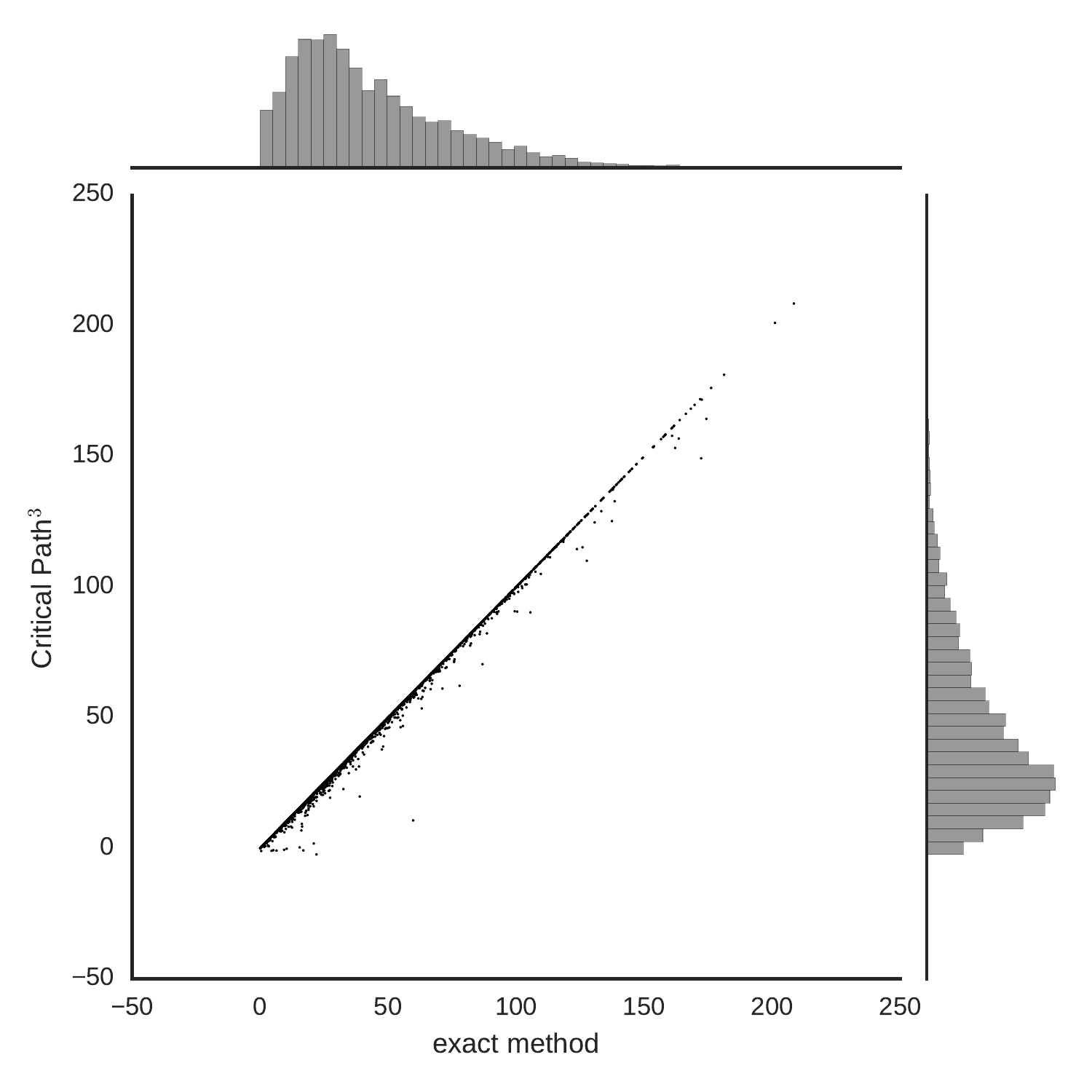}

  \caption{Left: Relative scores of the heuristics.  For each \textsc{Maximum
      Colorful Subtree} instance, we consider the relative score in
    comparison the exact method at 100\,\%.  For each method, instances are
    sorted with respect to relative score.  Right: Comparison of the score of
    the Critical Path$^3$ heuristic in comparison to the optimal score of the
    instance.  For each compound, we consider only the true molecular formula
    candidate.}
  \label{fig:scores}
\end{figure}
%%%%%%%%%%%%%%%%%%%%%%%%%%%%%%%%%%%%%%%%%%%%%%%%%%%%%%%%%%%%%%%%%%%%%%%%%%%%%

Third, we compared the scores reached by the different heuristics to the
scores of the exact solutions, see Fig.~\ref{fig:scores} (left).  For each
compound, we only considered the instance of the \textsc{Maximum Colorful
  Subtree} that corresponds to the true candidate molecular formula.  We
report scores relative to the exact solution (at 100\,\%), and sorted
instances with respect to this relative score.  In the resulting plot, it is
not obvious which of the heuristics ``Insertion'', ``Kruskal-style'' and
``Critical Path$^1$'' should be preferred.  We see that Critical Path$^3$ heuristic and, hence, the maximum of all methods are able to
compute (almost) optimal solutions for about 80\,\% of the instances.  In
turn, this means that even for these methods which perform excellent in
ranking the correct answer, we miss the optimal solution in about 20\,\% of
the instances.  In addition, we compared scores of the Critical Path$^3$
heuristic against the exact method in more detail, see Fig.~\ref{fig:scores}
(right): We see that for instances where the heuristic does not find the
optimal solution, the computed solution is only ``slightly suboptimal'' with
respect to its score.  In fact, Pearson correlation between the two measures
is $+1.00$.

%%%%%%%%%%%%%%%%%%%%%%%%%%%%%%%%%%%%%%%%%%%%%%%%%%%%%%%%%%%%%%%%%%%%%%%%%%%%%
\begin{figure}[tb]
  \centering
  \includegraphics[height=0.33\linewidth]{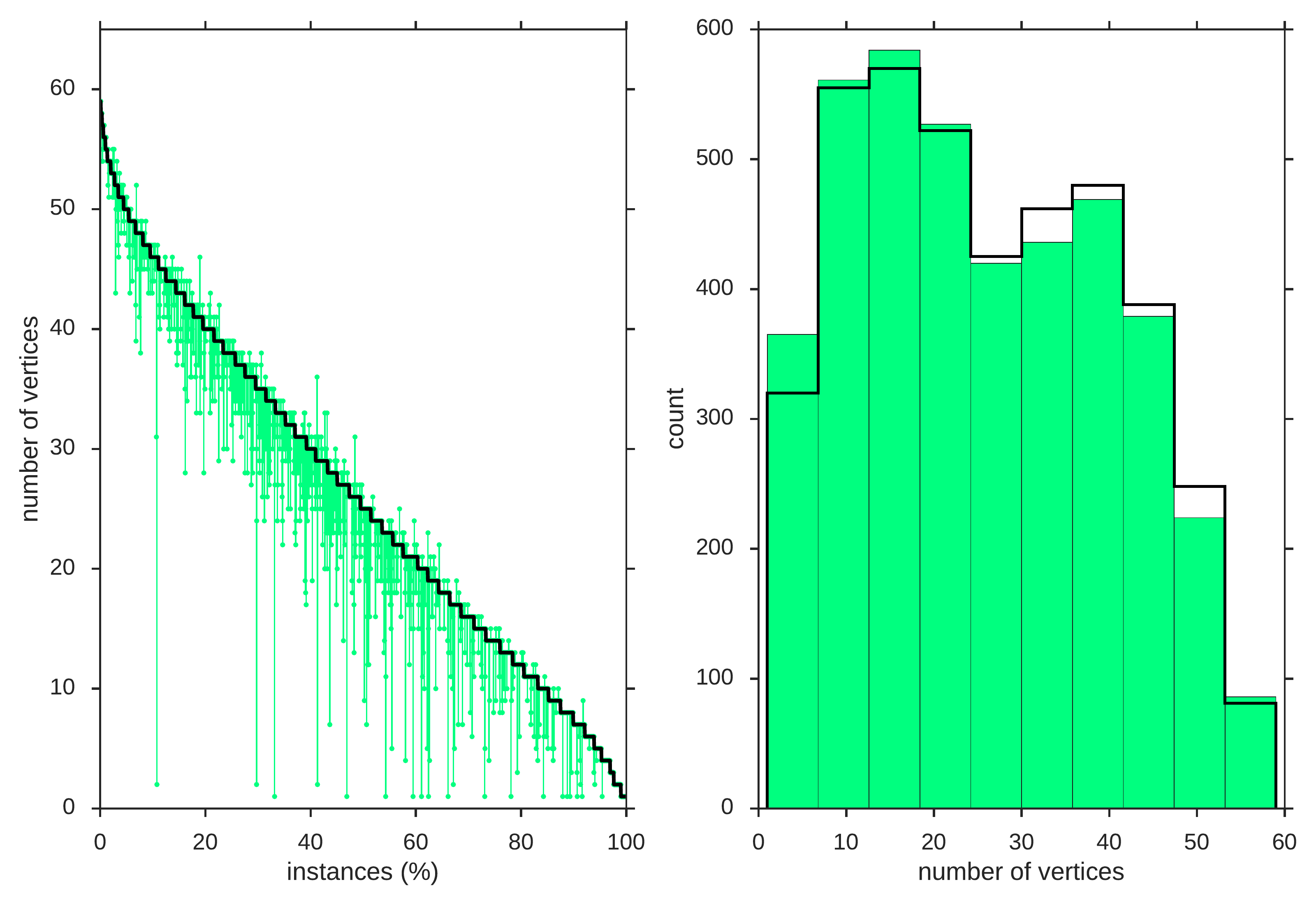}
  \hspace{0.02\linewidth}
  \includegraphics[height=0.33\linewidth]{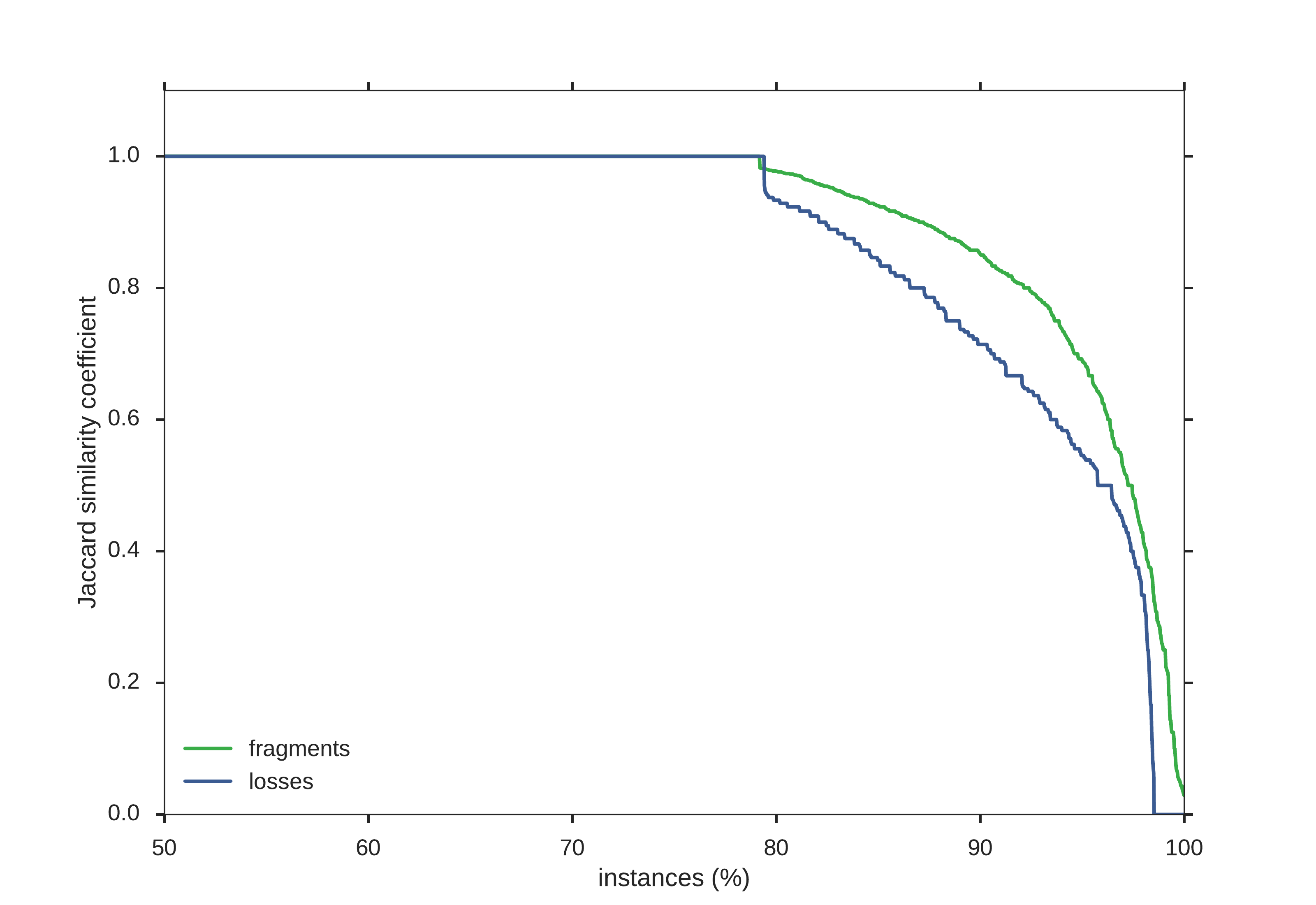}

  \caption{Left: Size of the fragmentation tree.  Instances are sorted with
    respect to the size of the optimal fragmentation tree (black); green bars
    indicate the corresponding tree sizes for the Critical Path$^3$
    heuristic.  Middle: Distribution of tree sizes for the exact method
    (black) and the Critical Path$^3$ heuristic (green).  Right: Comparison
    of the fragmentation tree structure, optimal tree vs.\ the tree computed
    by the Critical Path$^3$ heuristic.  Note the zoom of the x~axis.  In all
    three cases, we consider only the true molecular formula candidate for
    each compound.}
  \label{fig:structure}
\end{figure}
%%%%%%%%%%%%%%%%%%%%%%%%%%%%%%%%%%%%%%%%%%%%%%%%%%%%%%%%%%%%%%%%%%%%%%%%%%%%%

Fourth, we evaluate the solution structure quality of the Critical Path$^3$
heuristic.  Unfortunately, the ``true fragmentation tree'' cannot be
determined experimentally \cite{rasche11computing}.  To this end, we compare
heuristic tree structure vs.\ tree structures computed using the exact
method.  For each compound, we restrict the comparison to the true candidate
molecular formula; for other candidate molecular formulas, the optimal tree
cannot possibly be the ```true fragmentation tree''.  See
Fig.~\ref{fig:structure}.  For tree sizes, we observe rather large deviations
between heuristic and optimal trees; in contrast, the overall distribution of
tree sizes is highly similar.  But if we compare tree structures, we observe
much larger differences between the Critical Path$^3$ heuristic and the exact
method: We measure structural similarity comparing either the set of node
labels (fragments) or the set of edge labels (losses) of the two trees.  We
estimate the similarity of two (finite) sets $A,B$ using the \emph{Jaccard
  similarity coefficient} $J(A,B) = \abs{A \cap B} / \abs{A \cup B} \in
[0,1]$.  We observe that more than 20\,\% of the heuristic trees differ from
the corresponding optimal tree; for at least 10\,\%, this difference is
significant.

\begin{comment}
Finally, we showcase the speedup that can be expected in practice when
analyzing a complete LC-MS dataset.  \todo{Mach wie du magst, beschreib es
  grob hier.}
\end{comment}

%%%%%%%%%%%%%%%%%%%%%%%%%%%%%%%%%%%%%%%%%%%%%%%%%%%%%%%%%%%%%%%%%%%%%%%%%%%%%
%%%%  SECTION CONCLUDING REMARKS  %%%%%%%%%%%%%%%%%%%%%%%%%%%%%%%%%%%%%%%%%%%
%%%%%%%%%%%%%%%%%%%%%%%%%%%%%%%%%%%%%%%%%%%%%%%%%%%%%%%%%%%%%%%%%%%%%%%%%%%%%

\section{Conclusion}

We have presented heuristics for the \textsc{Maximum Colorful Subtree}
problem.  Our evaluation shows that the Critical Path$^3$ heuristic is
well-suited for choosing the correct candidate molecular formula, when applied
to tandem mass spectrometry data of small molecules.  Our evaluation
sidesteps the catch-22 that we want to evaluate solutions based on structure
and not score when, at the same time, the correct solution structure is not
known.  We have shown that the tree computed by the Critical Path$^3$
heuristic is often identical to the optimal tree.  Even when the heuristic
returns a suboptimal solution, the score is usually very close to the optimal
score.  In contrast, the structure of the heuristic tree deviates
significantly from the optimal tree for more than 20\,\% of the instances.
To this end, we argue not to use this tree for downstream analysis, such as
estimating chemical similarity based on fragmentation tree similarity
\cite{rasche12identifying} or machine learning \cite{shen14metabolite,
  duehrkop15searching, brouard16fast, brouard17magnitude-preserving}:
Preliminary evaluations clearly indicate that using trees computed by any
heuristic, leads to significantly worse results for the downstream analysis.
A back-of-the-envelope calculation indicates the problem: If we assume that
20\,\% of the heuristic trees are ``structurally faulty'', then a pairwise
comparison of trees will result in 36\,\% tree pairs where at least one of
the trees is ``structurally faulty''.

Building an instance DAG requires more time than running any of the presented
heuristics.  We conjecture that there is only limited potential for speeding
up the graph building phase.  To this end, whereas searching for better (and
not significantly slower) heuristics is still a valid undertaking, faster
heuristics are of little practical use.  It is worth mentioning that
computing exact solutions for the NP-hard \textsc{Maximum Colorful Subtree}
problem takes only about 100-fold the time needed for constructing the graph
instances; further speed-up is possible using data reductions and a stronger
ILP formulation of the problem from~\cite{white15speedy}.

Even elaborate heuristics for a bioinformatics problem, which are capable of
finding solutions with objective function value very close to the optimum,
can result in solutions which are structurally very dissimilar from the
optimum structure.  We showed that this is not only a theoretical
possibility, but happens regularly for real-world instances.  This underlines
the importance of finding exact solutions for bioinformatics problems; the
structure of solutions found by heuristic, including local search heuristics
such as Markov chain Monte Carlo, may deviate significantly from the optimal
solution.

%%%%%%%%%%%%%%%%%%%%%%%%%%%%%%%%%%%%%%%%%%%%%%%%%%%%%%%%%%%%%%%%%%%%%%%%%%%%%
%%%%  BIBLIOGRAPHY ETC  %%%%%%%%%%%%%%%%%%%%%%%%%%%%%%%%%%%%%%%%%%%%%%%%%%%%%
%%%%%%%%%%%%%%%%%%%%%%%%%%%%%%%%%%%%%%%%%%%%%%%%%%%%%%%%%%%%%%%%%%%%%%%%%%%%%

\paragraph{Acknowledgments.}

WTJW funded by Deutsche Forschungsgemeinschaft (grant BO~1910/9).

\paragraph{Availability.}

The Critical Path$^3$ heuristic and the exact method are available trough the
SIRIUS software (\url{https://bio.informatik.uni-jena.de/software/sirius/})
and also from GitHub (\url{https://github.com/boecker-lab/sirius}).  Source
code for all other heuristics will be made available upon request.  Instances
will be made available from our website.

\bibliographystyle{splncs03_mod}
\bibliography{bibtex/group-literature}

%%%%%%%%%%%%%%%%%%%%%%%%%%%%%%%%%%%%%%%%%%%%%%%%%%%%%%%%%%%%%%%%%%%%%%%%%%%%%

\end{document}